\documentclass[aps,prb,twocolumn]{revtex4}
\usepackage{amssymb}
\usepackage{graphicx}
\usepackage{hyperref}
\begin{document}
\title{Crystal growth and ambient and high pressure study of the reentrant superconductor Tm$_2$Fe$_3$Si$_5$}
\author{Yogesh Singh}\altaffiliation[Current Address:]{ Ames Laboratory and Department of Physics and Astronomy, Iowa State University, Ames, IA-50011, U.S.A.} \author{S. Ramakrishnan}

\affiliation{Tata Institute of Fundamental Research, Mumbai-400005, India}
\date{\today}

\begin{abstract}
Tm$_2$Fe$_3$Si$_5$ is known to undergo a transition to the superconducting state (at ambient or applied pressure depending on the sample) at a temperature $T_{\rm c1} ~(\sim 1.8$~K) and at a lower temperature $T_{\rm N} ~(\approx 1$~K) undergoes a transition into a long range antiferromagnetically ordered state.  The superconductivity is simultaneously destroyed and the sample reenters the normal state at $T_{\rm c2} = T_{\rm N}$\@.  The conditions reported in literature for the observation of superconductivity in Tm$_2$Fe$_3$Si$_5$ are sample dependent but it is now accepted that stoichiometric Tm$_2$Fe$_3$Si$_5$ superconducts only under pressure.    
Here we report single crystal growth of stoichiometric Tm$_2$Fe$_3$Si$_5$ which does not superconduct at ambient pressure down to 100~mK\@.  Measurements of the anisotropic static magnetic susceptibility $\chi(T)$ and isothermal magnetization $M(H)$, ac susceptibility $\chi_{\rm ac}(T)$, electrical resistivity $\rho(T)$ and heat capacity $C(T)$ at ambient pressure and $\chi_{\rm ac}(T)$ at high pressure are reported   The magnetic susceptibility along the $c$-axis $\chi_c(T)$ shows a small maximum around 250~K and does not follow the Curie-Weiss behavior while the magnetic susceptibility along the $a$-axis $\chi_a(T)$ follows a Curie-Weiss behavior between 130~K and 300~K with a Weiss temperature $\theta$ and an effective magnetic moment $\mu_{\rm eff}$ which depend on the temperature range of the fit.  The easy axis of magnetization is perpendicular to the $c$-axis and $\chi_{\rm a}$/$\chi_{\rm c}$~=~3.2 at 1.8~K\@.  The ambient pressure $\chi_{\rm ac}(T)$ and $C(T)$ measurements confirm bulk antiferromagnetic ordering at $T_{\rm N}$~=~1.1~K\@.  The sharp drop in $\chi_{\rm ac}(T)$ below the anti-ferromagnetic transition is suggestive of the existence of a spin-gap.  We observe superconductivity only under applied pressures $P\geq 2$~kbar\@.  The temperature-pressure phase diagram showing the non-monotonic dependence of the superconducting transition temperature $T_{\rm c}$ on pressure $P$ is presented.  
\end{abstract}  
\maketitle

\section{Introduction}
\label{sec:INTRO}
\noindent
The series of compounds R$_2$Fe$_3$Si$_5$ (R~=~ Sc, Y, Sm, and Gd--Lu) show interesting magnetic and superconducting properties.\cite{braun1980,braun1981,moodenbaugh1981,viningPhD,vining1983,vining1983a}  Compounds of this series containing magnetic rare-earth elements Gd--Er show antiferromagnetic ordering of trivalent rare-earth moments\cite{viningPhD,vining1983} and Yb$_2$Fe$_3$Si$_5$ has recently been shown to be a strongly correlated Kondo lattice system with a heavy Fermi liquid ground state surviving inside an antiferromagnetically ordered state.\cite{singh}  The non-magnetic compounds with R~=~Sc, Y, and Lu show superconductivity at relatively high temperatures ($T_{\rm c}$~=~4.4, 1.8 and 6.2~K respectively) given the large Fe content.\cite{vining1983a}  Moessbauer measurements on R$_2$Fe$_3$Si$_5$ have shown that there is no moment on the Fe and it only contributes in building a large density of states at the Fermi energy.\cite{braun1981, cashion1980,cashion1981} 	
The compound Tm$_2$Fe$_3$Si$_5$ is one of the most interesting in this series.  It was shown to be the first reentrant antiferromagnetic superconductor where superconductivity at $T_{\rm c1}$~$\sim$~1.5--1.8~K is destroyed by the onset of antiferromagnetism at $T_{\rm c2} = T_{\rm N}$ $\sim 1~K < T_{\rm c1}$\@.\cite{segre1981,vining1985, gotaas1987,moodenbaugh1985}  Tm$_2$Fe$_3$Si$_5$ therefore is a unique system unlike other stochiometric reentrant superconductors which either become superconducting again at low temperature once the antiferromagnetic order is established (such as, HoNi$_2$B$_2$C)\cite{Grigereit1994} or become normal due to ferromagnetism in the first place (ErRh$_4$B$_4$).\cite{fertig1977}  It is still not understood why the antiferromagnetic order among Tm$^{3+}$ ions is a deterrent to the superconductivity in Tm$_2$Fe$_3$Si$_5$ given that earlier neutron scattering studies \cite{gotaas1987} could not find any ferromagnetic moment below T$_N$ which could compete with the superconductivity.  

Although there have been several studies on this material, there are varying reports about the onset of superconductivity in Tm$_2$Fe$_3$Si$_5$ with some groups reporting the onset of superconductivity below 1.8~K at ambient pressure\cite{segre1981,moodenbaugh1985} while others observe superconductivity only under applied pressure.\cite{vining1985,gotaas1987}  It has been shown in a recent study that the homogeneity range of Tm$_2$Fe$_3$Si$_5$ is small and that any internal pressure due to the off-stoichiometry and the presence of impurity phases can lead to the varying values of $T_{\rm c1}$\@.\cite{schmidt1996}  That study also showed that stoichiometric Tm$_2$Fe$_3$Si$_5$ will superconduct only under pressure.\cite{schmidt1996}

A detailed experimental investigation of the magnetic, thermal and electrical properties of stoichiometric Tm$_2$Fe$_3$Si$_5$ has not been reported before.  In this work we report the growth of stoichiometric single crystalline samples of Tm$_2$Fe$_3$Si$_5$ and their electrical resistivity $\rho$ versus temperature $T$, static $\chi$ and ac magnetic susceptibility $\chi_{\rm ac}$ versus $T$, magnetization $M$ versus magnetic field $H$, and heat capacity $C$ versus $T$ at ambient pressure and $\chi_{\rm ac}$ versus $T$ at high pressure.  Our data suggest valence fluctuations and/or the partial gapping of the Fermi surface below the anti-ferromagnetic ordering temperature as possible mechanisms for the destruction of superconductivity at low temperatures. 
 

\section{EXPERIMENTAL DETAILS}
\label{sec:EXPT}
\noindent
Single crystals of Tm$_2$Fe$_3$Si$_5$ were grown by a modified vertical Bridgeman method.  The purity of the rare-earth element Tm and the transition-metal element Fe were 99.99\% whereas the purity of Si was 99.999\%\@.  First, 5 grams of a polycrystalline sample was prepared by arc-melting the constituent elements.  The Fe and Si were taken in stoichiometric proportions and about 5\% extra Tm over the stoichiometric amount was taken to compensate for any loss of Tm, due to its high vapor pressure, during the synthesis.  The resulting tablet was remelted several times to promote homogeneity.  The mass loss at this stage was negligible.  This polycrystalline tablet was crushed into a fine powder and placed in an Alumina crucible.  This in turn is placed inside a tantalum container and sealed in an arc furnace under ultra high purity argon atmosphere.  The whole assembly is placed in a vertical tube furnace and slowly heated to 1450~$^\circ$C in a dynamic Ar atmosphere and kept there for 4--6~hours.  The sample assembly is then pulled out of the furnace at a rate of 2--3~mm/hr\@.  

The temperature dependence of the magnetic susceptibility $\chi(T)$ and the isothermal magnetization $M(H)$ with the magnetic field along the $a$, $b$ or $c$-axes, was measured using the horizontal rotator option on the commercial SQUID magnetometer (MPMS5, Quantum Design).  The resistivity $\rho(T)$ between 1.5~K and 300~K was measured using an LR-700 ac resistance bridge by the 4-probe method with contacts made by silver paste.  The heat-capacity $C(T)$ in zero field between 0.3 to 35~K was measured using the commercial PPMS (Quantum Design).  The ambient pressure ac susceptibility $\chi_{ac}(T)$ between 0.1~K and 4~K was measured at a frequency of 1~kHz, with an ac amplitude of 1~Gauss, using an adiabatic demagnetization fridge (Cambridge Magnetic Refrigeration, Cambridge).  The high pressure $\chi_{ac}(T)$ between 1.8~K and 10~K were measured using the mutual induction option of the LR-700 ac resistance bridge.  In this case the primary and the secondary coils, each of roughly 100 turns of 0.25~mm Cu wire, is wound around a cylinder of thin Mylar sheet, with the sample inside the cylinder, and then placed inside a piston-clamp type Cu-Be high pressure cell with Daphne oil as the pressure transmitting medium.  The pressure so generated is hydrostatic.  The pressure at low temperatures was determined by the superconducting transition of a small ($\sim$~1~mg) piece of lead Pb which was loaded in the cell with the sample.

\section{RESULTS}
\subsection {Crystal Growth and Structure}
\begin{figure}[t]
\includegraphics[width=3.in]{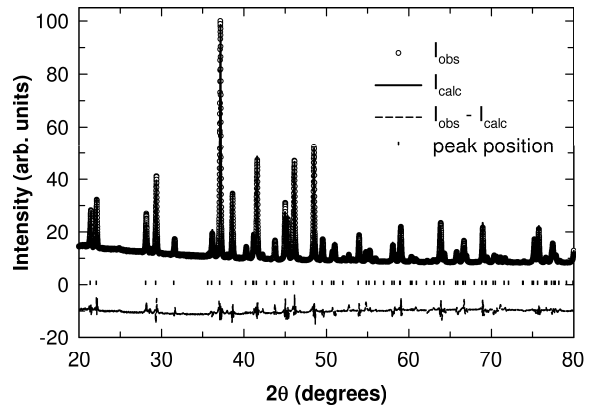}
\caption{ Rietveld refinement of the Tm$_2$Fe$_3$Si$_5$ powder X-ray diffraction data. The open symbols represent the observed data, the solid lines through the data represent the fitted pattern, the dotted lines represent the difference between the observed and calculated intensities and the vertical bars represent the peak positions.
\label{figxrd}}
\end{figure}

The tantalum container was cut open using a saw and the Al$_2$O$_3$ crucible removed.  Inside this crucible one could see a melted mass of material stuck to the crucible walls.  The Al$_2$O$_3$ crucible was cut off from the bottom and the sides using a diamond-wheel-cutter.  The cutting is done carefully to minimize the loss of material.  The melted ingot is then lightly tapped from the sides with a hammer and crystals in the form of platelets with a largest size of 1.25$\times$1$\times$.25~mm$^3$ are physically extracted.  Some of the smaller crystals were crushed for powder X-ray diffraction (XRD).  Cobalt radiation (Co K$\alpha$) was used to suppress the flouresence from the Fe.  The XRD of the crushed crystals (shown in Fig.~\ref{figxrd}) confirmed the structure and absence of impurity phases and sharp Laue back scattering patterns indicated that the crystals were of high quality.  All the peaks in the powder XRD pattern could be indexed to the known Sc$_2$Fe$_3$Si$_5$-type structure (space group $P4/mnc$) and a Reitveld refinement, shown as the solid curve through the data in Fig.~\ref{figxrd}, gave the lattice constants $a$~=~$b$~=~10.371(3)~\AA\ and $c$~=~5.403(5)~\AA\ which are in good agreement with the literature value ($a$~=~$b$~=~10.367~\AA\ and $c$~=~5.407~\AA\ ).\cite{schmidt1996}  From Laue back scattering pictures the largest surface of the plate-like crystals were found to be perpendicular to either the \emph{a}-axis or the \emph{b}-axis.  The Laue measurements were also used to find the (001) orientation (the tetragonal $c$-axis) of the crystals.

\subsection{Ambient Pressure Measurements}
\subsubsection{Magnetic Susceptibility }
\begin{figure}[t]
\includegraphics[width=3.in]{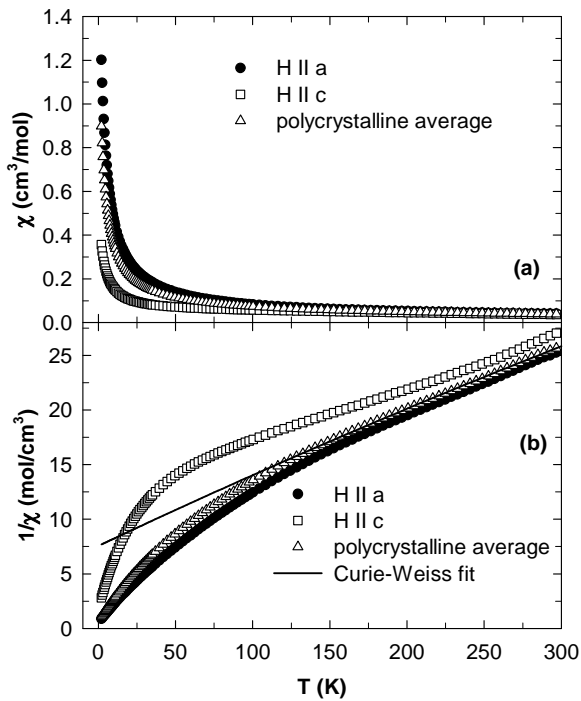}
\caption{(a) Temperature $T$ dependence of the static magnetic susceptibility $\chi$ of Tm$_2$Fe$_3$Si$_5$ from 1.8 to 300~K in a field of 1~kOe applied along the $a$- and $c$-axes.  The polycrystalline average susceptibility is also shown.  (b) Temperature $T$ dependence of the inverse susceptibility (1/$\chi$).  The solid line is a fit to the Curie-Weiss relation (see text). 
\label{figsus1}}
\end{figure}

In Fig.~\ref{figsus1} we present the results of the anisotropic static susceptibility $\chi$ measurements.  Figure~~\ref{figsus1}(a) shows the $\chi (T)$ data between 1.8 and 300~K in a field of 1~kOe applied along the $a$- and $c$-axes.  As expected for a tetragonal system, the susceptibility along the $a$- and $b$-axes (not shown) is the same.  The polycrystalline averaged susceptibility $\chi_{\rm poly}$ which can be calculated as 
\begin{equation}
\chi_{\rm poly}(T)~=~{\chi_{a}(T)+\chi_b(T)+\chi_c(T)\over 3}~,
\label{eqpoly}
\end{equation}
is also shown.  
\begin{figure}[t]
\includegraphics[width=3.in]{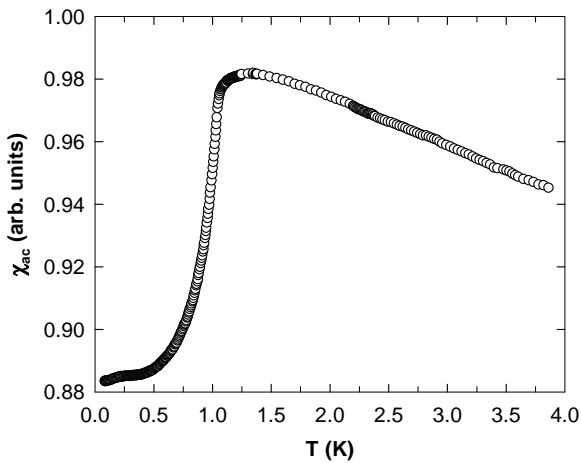}
\caption{Temperature dependence  of the zero field ac susceptibility $\chi_{ac}$ of Tm$_2$Fe$_3$Si$_5$ from 100~mK to 4~K. The sharp anomaly peaked around 1.15~K is the signature of antiferromagnetic ordering.
\label{figsus2}}
\end{figure}

The susceptibility is moderately anisotropic with the easy axis of magnetization being perpendicular to the tetragonal $c$-axis and $\chi_{a}$/$\chi_c$~=~3.2 at 1.8~K\@.  Within a series of iso-structural rare-earth based compounds, the magneto-crystalline anisotropy changes from uniaxial (easy axis along $c$-axis) to easy plane (easy axis perpendicular to $c$-axis) behavior at Er.\cite{buschow2003}  These systematics have been observed in $\chi(T)$ measurements on oriented polycrystals of other magnetic members ($R$~=~Gd--Er) of this series.\cite{braun1981}  The results for Tm$_2$Fe$_3$Si$_5$ therefore, follow the expected trend.  

The plots of the inverse susceptibilities are shown in Fig.~\ref{figsus1}(b).  
The 1/$\chi_{\rm c}$ data in Fig.~\ref{figsus1}(b) shows a curvature in the whole temperature range with a broad minimum at about 250~K and could not be fitted by a Curie-Weiss expression.  
The 1/$\chi_{\rm a}$ data between 130~K and 300~K could be fitted by a modified Curie-Weiss expression given by,
\begin{equation}
 \chi~=~\chi_0~+~{C \over (T~-~\theta)}~. 
\label{eqCW}
\end{equation}
The fit (not shown here) gave the values $\chi_0$~=~4.3(8)$\times 10^{-3}$~cm$^3$/Tm mol, $C$~=~7.0(3)~cm$^3$/Tm~mol~K, and $\theta$~=~-88(5)~K\@.  The value $C$~=~7.0(3)~cm$^3$/Tm~mol~K corresponds to an effective moment of $\mu_{\rm eff}$~=~7.5(2)$\mu_B$ which is close to the free moment value 7.56$\mu_B$ for trivalent Tm moments.  However, fitting the 1/$\chi_{\rm a}(T)$ data between 300~K and some temperature $T^*$ above 130~K, gave $\theta$ and $C$ values which depended on $T^*$\@.  For example, with $T^*$~=~200~K, the fit gave $\chi_0$~=~3.9(5)$\times 10^{-3}$~cm$^3$/Tm mol, $C$~=~6.7(1)~cm$^3$/Tm~mol~K, and $\theta$~=~-65(3)~K\@.   
The 1/$\chi_{\rm poly}(T)$ data between 150~K and 300~K could also be fit by Eq.~(\ref{eqCW}).  The fit, shown as the solid line extrapolated to $T$~=~0~K, in Fig.~\ref{figsus1}(b), gave the values $\chi_0$~=~4.0(4)$\times 10^{-3}$~cm$^3$/Tm mol, $C$~=~6.9(3)~cm$^3$/Tm~mol~K, and $\theta$~=~-109(6)~K.  The value $C$~=~6.9(3)~cm$^3$/Tm~mol~K corresponds to an effective moment of $\mu_{\rm eff}$~=~7.4(4)$\mu_B$\@.  The dependence of the fitting parameters on the temperature range of the fit was also found for the $\chi_{\rm poly}$ data.\\
To the best of our knowledge there is only one previous report of the static magnetic susceptibility on a polycrystalline sample of this material.\cite{schmidt1996}  It was reported in Ref.~\onlinecite{schmidt1996} that the magnetic susceptibility data between 70~K and 300~K could be fitted by a Curie-Weiss law plus a constant term with an effective moment 7.0(3)~$\mu_{\rm B}$\@.  Neither the value of $\theta$ was reported, nor the effect of changing the temperature range of the fit was discussed.  Comparing our 1/$\chi(T)$ data in Fig.~\ref{figsus1}(b) to those of the polycrystalline sample in Fig.~7 of Ref.~\onlinecite{schmidt1996} it is clear that the polycrystalline sample must have had a large preferential orientation along the $c$-axis.  A definite curvature in the 1/$\chi(T)$ data at high temperatures of the polycrystalline sample in Fig.~7 of Ref.~\onlinecite{schmidt1996} can be seen.
  
The large value of $\theta$ compared to the magnetic ordering temperature (discussed below) and the temperature dependence of the $\chi(T)$ data above are interesting and we will return to these in Sec.~\ref{sec:Dis} below when we discuss our results.

Figure~\ref{figsus2} shows the ambient pressure ac susceptibility $\chi_{\rm ac}$ data between 0.1~K and 4~K in zero applied dc magnetic field.  The peak in $\chi_{\rm ac}$ at $T_{\rm N}$~=~1.15~K indicates the onset of antiferromagnetic ordering.  The shape of the anomaly at $T_{\rm N}$ is unusual and different from what is usually observed at an antiferromagnetic transition.  The sharp drop in $\chi_{\rm ac}$ by more than 10\% and the weak temperature dependence at lower temperatures suggests the presence of a spin gap in this system below $T_{\rm N}$\@.  We observe a small bump in $\chi_{\rm ac}$ below 0.25~K the origin of which is not clear at present.
We did not observe any signature of superconductivity down to 0.1~K\@.

\subsubsection{Magnetization}
\begin{figure}[t]
\includegraphics[width=3.in]{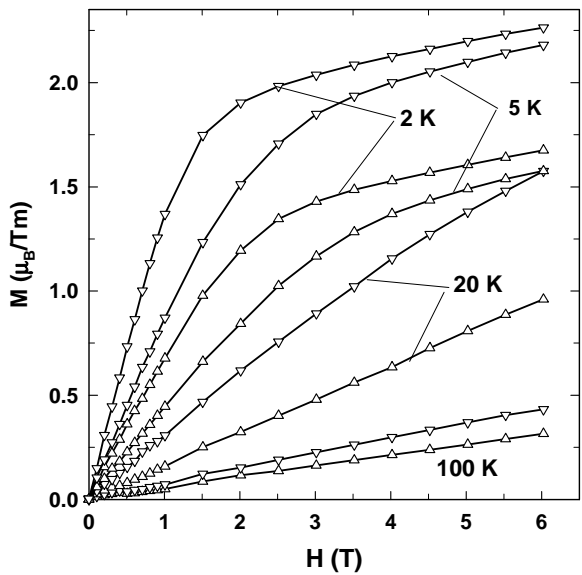}
\caption{Anisotropic isothermal magnetization $M$ of Tm$_2$Fe$_3$Si$_5$ versus magnetic field $H$ at various  temperatures, with $H$ along the $a$-axis ($\nabla$) and $c$-axis ($\Delta$).
\label{figsus3}}
\end{figure}

The magnetization $M$ versus field $H$ at various temperatures $T$ is shown in  Fig.~\ref{figsus3}.  At high temperatures the $M$ data are linear with applied field and the anisotropy between the $a$- and $c$-axis magnetization is small.  The $M(H)$ data begin to show a curvature below 20~K and in the data for 5~K and 2~K it can clearly be seen that the anisotropy between the $a$- and $c$-axes is larger and there is a tendency of saturation on increasing the field.  The magnetization however, continues to increase with field up to the highest applied field, $H$~=~6~T\@.  The magnetization at 2~K at the highest field 6~T is much smaller than the expected value $gJ$~=~12~$\mu_{\rm B}$ for Tm$^{3+}$ moments. 

\subsubsection{Resistivity}
\begin{figure}[t]
\includegraphics[width=3.in]{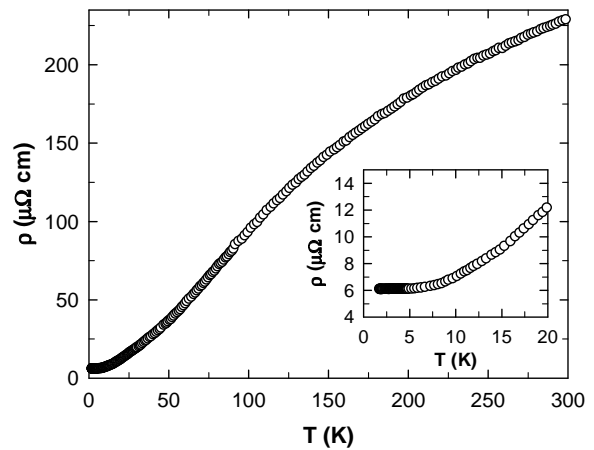}
\caption{Resistivity $\rho$ versus temperatures $T$ of Tm$_2$Fe$_3$Si$_5$ with current $I$ along the $a$-axis.
\label{figres}}
\end{figure}

The resistivity $\rho(T)$ of Tm$_2$Fe$_3$Si$_5$ between 1.5~K and 300~K for excitation current $I$ along the $a$-axis is shown in the main panel of Fig.~\ref{figres}. The sample exhibits normal metallic behavior with a room temperature value $\rho$(300~K)~=~233~$\mu \Omega$~cm\@.  The inset shows the low temperature data between 1.5~K and 20~K on an expanded scale.  The small value of the residual resistivity $\rho(1.5~K)$~=~6~$\mu \Omega$~cm and a reasonably large residual resistivity ratio RRR~=~$\rho(300~K)/\rho(1.5~K)$~=~39 indicates good sample quality.  We did not observe any signature of superconductivity down to 1.5~K in our resistivity measurements.  The temperature dependence of $\rho$ is qualitatively similar to the only other resistance measurement that was reported on a polycrystalline sample which showed a partial superconducting transition below 1.8~K at ambient pressure.\cite{segre1981}  The resistivity was not reported and therefore we can not compare those data with our resistivity values.  The residual resistivity ratio for the polycrystal was about $40$ which is close to what we observe for our single crystal sample.

\subsubsection{Heat Capacity}
\begin{figure}[t]
\includegraphics[width=3.in]{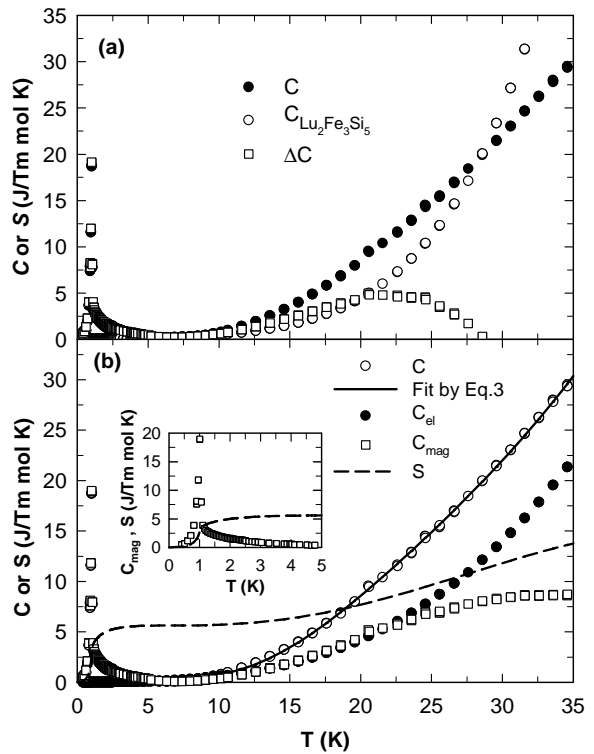}
\caption{(a) The heat capacity $C$ versus temperature $T$ of Tm$_2$Fe$_3$Si$_5$ from 0.3 to 35~K. Also shown are the heat capacity of Lu$_2$Fe$_3$Si$_5$ $C_{\rm Lu_2Fe_3Si_5}$ and the difference $\Delta C$ between $C$ and $C_{\rm Lu_2Fe_3Si_5}$.  (b) The heat capacity $C$ versus $T$, the fit by Eq.~(\ref{Eq3}), the magnetic heat capacity C$_{\rm mag}$, and the entropy $S$ versus $T$.  The inset in (b) shows the low temperature  C$_{\rm mag}$ and $S$ data between 0.3 and 5~K\@. 
\label{figHC}}
\end{figure}

The temperature dependence of the heat capacity $C$ of Tm$_2$Fe$_3$Si$_5$ between 0.3 and 35~K is shown in Fig.~\ref{figHC}(a).  The sharp peak of nearly 20~J/Tm~mol~K just above T$_{\rm N}$~=~1~K confirms the bulk antiferromagnetic ordering in this compound.  To estimate the magnetic contribution to $C$ we tried to use the heat capacity $C_{\rm Lu_2Fe_3Si_5}$ of the isostructural nonmagnetic compound Lu$_2$Fe$_3$Si$_5$.  The $C_{\rm Lu_2Fe_3Si_5}$ data, shown in Fig.~\ref{figHC}(a) was taken from our previous paper.\cite{singh}  The difference $\Delta C$ between $C$ and $C_{\rm Lu_2Fe_3Si_5}$ is also shown in Fig.~\ref{figHC}(a).  The $C_{\rm Lu_2Fe_3Si_5}$ is larger than $C$ above $T$~=~28~K which gives an unphysical negative $\Delta C$ as seen in Fig.~\ref{figHC}(a).  This indicates that the lattice dynamics of the two materials are probably different and that $C_{\rm Lu_2Fe_3Si_5}$ is not an appropriate estimation of the lattice heat capacity for Tm$_2$Fe$_3$Si$_5$.  The $\Delta C(T)$ data in Fig.~\ref{figHC}(a) suggests that there is a contribution to the $C$ from a Schottky like anomaly above 20~K\@.  To estimate the electronic $C_{\rm el}$, lattice $C_{\rm ph}$ and Schottky $C_{\rm Schottky}$ contributions to $C$ we fitted the $C(T)$ data between 10~K and 35~K using the expression\cite{Kittel}

\begin{eqnarray}
C &=& C_{\rm el} + C_{\rm ph} + C_{\rm Schottky} \nonumber\\
&=& \gamma T + \beta T^3 + {g_0\over g_1}\Big({\delta \over T}\Big)^2 {\exp({\delta\over T})\over \Big[1 + {g_0\over g_1}\exp({\delta\over T})\Big]^2}~~,
\label{Eq3}	
\end{eqnarray}
where, $\gamma$ is the Sommerfeld coefficient, $\delta$ is the energy splitting between a ground state of degeneracy $g_0$ and an excited state of degeneracy $g_1$.  The fit, shown as the solid curve through the $C(T)$ data in Fig.~\ref{figHC}(b), gave the values $\gamma$~=~56(4)~mJ/Tm~mol~K$^2$, $\beta$~=~0.94(4)~mJ/Tm mol~K$^4$, $g_0/g_1$~=~.329(3), and $\delta$~=~98(3)~K\@.  From the value of $\beta$ one can estimate the Debye temperature $\theta_{\rm D}$ using the expression\cite{Kittel}  
\begin{equation}
\Theta_{\rm D}~=~\bigg({12\pi^4{\rm R} n \over 5\beta}\bigg)^{1/3}~, 
\label{EqDebyetemp}
\end{equation}
\noindent
where, R is the molar gas constant and $n$ is the number of atoms per formula unit ($n$~=~10 for Tm$_2$Fe$_3$Si$_5$).  We obtain $\theta_{\rm D}$~=~268(8)~K\@.

The magnetic contribution C$_{mag}$ which was obtained by subtracting $C_{\rm el}$ and $C_{\rm ph}$ from $C$, and the estimated entropy $S$ obtained by integrating the $C_{\rm mag}$/$T$ versus $T$ data are also shown in Fig.~\ref{figHC}~(b).   The inset in Fig.~\ref{figHC}~(b) shows $C_{\rm mag}$ and $S$ between 0.3 and 5~K to highlight the low temperature behavior.  The entropy reaches a value of 4.53~J/Tm~mol~K at $T_{\rm N}$ which is smaller than the value R~ln2~=~5.76~J/Tm~mol~K expected for a doublet ground state.  An entropy of R~ln2 is reached only near 5~K after which the entropy increases only weakly up to about 10~K\@.  The tail in $C(T)$ above $T_{\rm N}$ and the weak temperature dependence of $S$ between 5~K and 10~K strongly indicate that the ground state is a quasi-doublet with a small splitting due to the crystalline electric field (CEF).  For higher temperature the entropy increases again probably due to contributions from a Shottky-like anomaly as indicated in the $C_{\rm mag}(T)$ data in Fig.~\ref{figHC}~(b)\@.  The value of $\delta$ obtained above indicates that the first excited CEF level is about 100~K above the quasi-doublet ground state.  
    
At 35~K the entropy reaches a value of about 13~J/Tm~mol~K which is much smaller compared to the full R~ln(2J+1) ($\approx$~22~J/Tm~mol~K) expected for Tm$^{3+}$ (J~=~6) moments.  Considerably higher temperatures need to be attained to observe the full entropy. 

\subsection{High Pressure ac Susceptibility}

\begin{figure}[t]
\includegraphics[width=3.in]{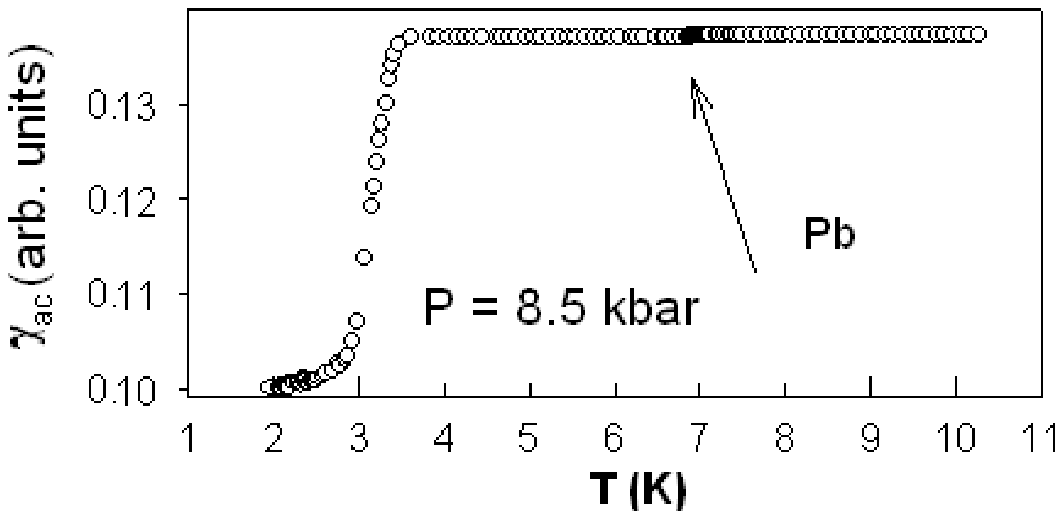}
\caption{Temperature $T$ dependence of $\chi_{\rm ac}$ of Tm$_2$Fe$_3$Si$_5$ under a pressure of 8.5~kbar\@.   The pressure was determined by the superconducting transition of lead (Pb) indicated by the arrow. 
\label{figHPsus}}
\end{figure}

The temperature dependence of the ac susceptibility $\chi_{\rm ac}(T)$ of Tm$_2$Fe$_3$Si$_5$ was measured between 1.8~K and 10~K with various applied pressures $P$\@.  There is no superconducting transition at ambient pressure.  We observe the onset of a superconducting transition below 1.9~K (not shown) only at a pressure of about 2~kbar\@.  For higher pressures the superconducting transition temperature increases.  The $P$~=~8.5~kbar $\chi_{\rm ac}(T)$ data, for which we observe the highest superconducting transition, is shown in Fig.~\ref{figHPsus}.  The superconducting transition for a small piece of lead Pb loaded along with the sample is indicated by the arrow in the figure.  The abrupt diamagnetic signal below 3.1~K is the superconducting transition of Tm$_2$Fe$_3$Si$_5$\@.  

\begin{figure}[t]
\includegraphics[width=3.in]{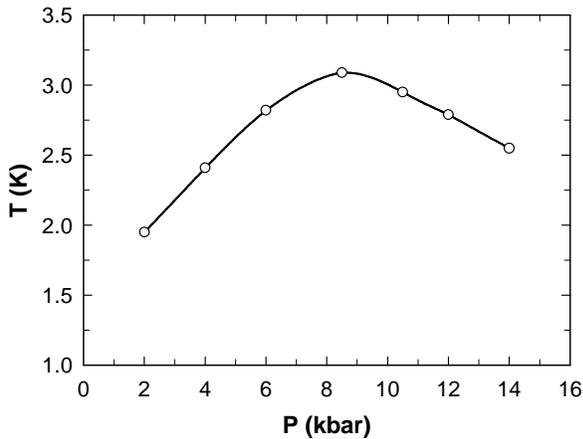}
\caption{Temperature $T$ versus Pressure $P$ phase diagram of Tm$_2$Fe$_3$Si$_5$ showing the pressure dependence of the superconducting transition temperature.  The line through the data is a guide to the eye.
\label{figPTphasediagram}}
\end{figure}

The temperature $T$ versus pressure $P$ phase diagram for Tm$_2$Fe$_3$Si$_5$ is shown in Fig.~\ref{figPTphasediagram}.  The non-linear dependence of $T_{\rm c}$ on pressure $P$ with a maximum at around 8.5~kbar is similar to the behavior reported for polycrystalline samples which do not show a superconducting transition at ambient pressure.\cite{schmidt1996}  The initial pressure coefficient $dT_{\rm c}/dP$ for our single crystalline sample was determined to be about 0.22~K/kbar\@.

\section{DISCUSSION}
\label{sec:Dis}
To start with, we did not observe superconductivity for our single crystalline sample at ambient pressure for temperatures down to 100~mK\@.  This indicates that our sample is close to the correct stoichiometry and is free of internal strains which could be present in the sample due to the off stoichiometry and/or the presence of impurity phases and in turn lead to ambient pressure superconductivity.\cite{schmidt1996}  Some of the earliest reports on this material suggested ambient pressure superconductivity and were therefore done on materials with either impurity phases and internal chemical pressure or materials which were off-stoichiometric.\cite{segre1981,moodenbaugh1985}    

In previous reports the magnetic susceptibility at the antiferromagnetic transition was also obscured by the fact that it occurs so close to the superconducting transition.  We have shown by measurements on the stoichiometric material that at $T_{\rm N}$ the shape of $\chi_{\rm ac}$ is unusual and it shows a sudden decrease below the transition before leveling off at a finite value at the lowest temperatures.  The sudden drop in $\chi$ below the transition and a finite non-zero susceptibility at the lowest temperatures probably indicate partial gapping of the Fermi surface below $T_{\rm N}$\@.  A spin-gap could occur if the magnetic order below $T_{\rm N}$ is accompanied by a spin density wave (SDW).  Neutron diffraction measurements, preferably on a stochiometric sample of Tm$_2$Fe$_3$Si$_5$ will be needed to check this possibility.  Resistivity measurements on a stochiometric sample down to temperatures below $T_{\rm N}$ would also be useful to look for signatures of partial gapping of the Fermi surface below $T_{\rm N}$\@.

The values of $\theta \sim -$90~K, obtained by fitting the $\chi_{\rm a}$ or the $\chi_{\rm poly}$ data are unusually large considering that antiferromagnetic ordering occurs at 1.1~K in this compound.  A large $\theta$ and a comparatively much smaller ordering temperature is sometimes seen in low dimensional systems or in systems where the magnetic interactions are frustrated.  The crystal structure of Tm$_2$Fe$_3$Si$_5$ does not suggest that either of these possibilities hold for this material.  Another class of systems where a large $\theta$ is seen are systems with valence fluctuations.\cite{batlogg1979}  Indeed the temperature dependence of the $\chi_{\rm c}(T)$ data with a maximum around 250~K is similar to what is observed for valence fluctuating materials.  The dependence of the values of the effective moment and the Weiss temperature on the temperature range of the Curie-Weiss fit is also what has been theoretically predicted for valence fluctuating Tm compounds.\cite{Aligia1984}  Valence-fluctuations and hybridization between the 4$f$ and conduction electrons are also supported by the moderately enhanced $\gamma$~=~56(4)~mJ/Tm~mol~K$^2$ in the paramagnetic state.
Valence fluctuations are known to suppress superconductivity and the application of pressure, which will stablize the Tm$^{3+}$ state and suppresse the valence fluctuations, will therefore enhance the superconductivity.  This is exactly what is observed for Tm$_2$Fe$_3$Si$_5$\@.  Our data therefore suggest that valence fluctuations of Tm moments and/or the partial gapping of the Fermi surface below $T_{\rm N}$ could be possible mechanisms for the destruction of superconductivity in Tm$_2$Fe$_3$Si$_5$\@.    

\section{CONCLUSION}
\label{sec:CON}
We have grown high quality, stoichiometric single crystals of the compound Tm$_2$Fe$_3$Si$_5$ using a modified bridgeman method and characterized them by X-ray, Laue back scattering, ac and static magnetic susceptibility, isothermal magnetization, resistivity, heat capacity and high pressure ac susceptibility measurements. Anomalies in the ambient pressure ac susceptibility and heat capacity confirm bulk antiferromagnetic ordering at $T_{\rm N}$~=~1.1~K in this compound.  The $\chi_{\rm ac}$ drops abruptly below $T_{\rm N}$ suggesting a partial gapping of the Fermi surface. The $\chi_{\rm a}$ and $\chi_{\rm poly}$ follow a Curie-Weiss behavior above 130~K with values of the Weiss temperature $\theta$ and the effective moment $\mu_{\rm eff}$ which depend on the temperature range of the fit.  The estimated $\mu_{\rm eff}$ comes out to be close to the value for a free trivalent Tm moment. The $\theta_P \sim -90~K$ is anomalously large given that T$_N$ for this compound is only 1.1~K.  These results suggest that valence fluctuations and/or gapping of the Fermi surface below $T_{\rm N}$ coul;d be possible mechanisms for the destruction of superconductivity in Tm$_2$Fe$_3$Si$_5$\@.  We did not observe any superconductivity in our ambient pressure measurements down to 100~mK\@.  Under an external pressure $P\geq 2$~kbar superconductivity is observed and a maximum $T_{\rm c}$~=~3.05~K is observed at a pressure of $P$~=~8.5~kbar.  We confirmed the non-monotonic dependence of the $T_{\rm c}$ on $P$\@.

\begin{acknowledgments}

YS would like to thank C. Geibel for useful discussions about the crystal growth and the Max Planck institute for the Chemical Physics of Solids, Dresden, for support during his visit in 2000-2001. 

\end{acknowledgments}

\end{document}